# LEVEL REPULSION IN INTEGRABLE SYSTEMS

Tao Ma and R. A. Serota

Department of Physics

University of Cincinnati

Cincinnati, OH 45244-0011

serota@ucmail.uc.edu

**Abstract**

Contrary to conventional wisdom, level repulsion in semiclassical spectrum is not just a feature of classically chaotic systems, but classically integrable systems as well. While in chaotic systems level repulsion develops on a scale of the mean level spacing, regardless of location in the spectrum, in integrable systems it develops on a much longer scale – geometric mean of the mean level spacing and the running energy in the spectrum. We show that at this scale level correlations in integrable systems have a universal dependence on level separation, as well as discuss their exact form at any scale. These correlations have dramatic consequences, including deviations from Poissonian statistics in the nearest level spacing distribution and persistent oscillations of level number variance as a function of the interval width. We illustrate our findings on two models – a rectangular infinite well and a modified Kepler problem – that serve as generic types of hard-wall billiards and potential problems. Our theory is based on the concept of parametric averaging, which allows for a statistical ensemble of integrable systems at a given spectral location (running energy).

# I. Introduction

Properties of semiclassical spectra have been subject of intense scrutiny over the past half-century. The two cases – of classically integrable systems without extra symmetries (degeneracy) and classically chaotic, ergodic systems – received the most attention. [1],[2] Chaotic systems are characterized by the Wigner-Dyson statistics of nearest level spacing and a weak logarithmic dependence of the level rigidity and level number variance on the width of the interval. Integrable systems were believed to exhibit Poissonian statistics of level spacing and saturation of the level rigidity and level number variance.[3]-[6]

Difficulty with the chaotic systems lies mainly in the fact that it is challenging to establish the relationship between the properties of the spectrum and underlying classical dynamics. Periodic orbit theory only recently developed some traction [7] outside the perturbative regime.[8] Exact description of spectral correlations was due to the random matrix theory (RMT) [1],[2] and the non-linear sigma model,[9] which produced identical results. The former assumes averaging over the spectrum while the latter is based on the Anderson model of electron diffusion, believed to be equivalent to a generic chaotic, ergodic motion. Furthermore, the latter provides an alternative approach to ensemble averaging – at a given energy in the spectrum but for various realizations of disorder (disorder averaging).[10]

Conversely, we shall argue that spectral properties of integrable systems are amenable to both semiclassical (based on classical dynamics and the periodic orbit theory) and direct quantum-mechanical derivations. Conclusions of our derivation challenge the commonly held belief of the absence of nearest level correlations and Poissonian spacing statistics. We show that a repulsive term in the level correlation function leads to a small but distinctive deviation from the latter. In retrospect, this should come as no surprise since it was established, numerically [3] and analytically [4], that the level rigidity and, presumably, the level number variance [6] achieved saturation for interval widths beyond a value proportional to the geometric mean of mean spacing and the running position in the spectrum. (In reality, the level number variance exhibits a far more complex behavior, as will be explained below.) This fact clearly contradicts the Poissonian absence of correlations, as the latter would imply an uninterrupted linear growth of those quantities.

Existence of a repulsive term in the level correlation function of integrable systems was established in our previous works, [11],[12] as was the fact that the level number variance exhibits persistent oscillations, as a function of the interval width, around what had been previously presumed to be its saturation value. That same repulsive term is responsible for the deviation from the Poissonian statistics of the nearest level spacing.

This paper is organized as follows. We first discuss the general properties of the level correlation function and its specific forms for the rectangular billiard and the modified Kepler problem. We proceed to derive the simplest form of the nearest level spacing distribution to account for the repulsive term in the level correlation function. We show that our numerical results, obtained on the basis of the concept of parametric ensemble averaging, are in excellent agreement with our prediction for the deviation from the Poissonian statistics.

## II. Level correlations in integrable systems.

In what follows, it is assumed that the spectrum is flattened [2], so that the mean level density $\langle \rho \rangle$ and the mean level spacing $\Delta$ are constant (and are set to unity – that is, all energies are measured in the units of $\Delta$):

$$\Delta = \langle \rho \rangle^{-1} = 1 \quad (1)$$

Deviation form the mean, $\delta\rho(\varepsilon) = \rho(\varepsilon) - \langle \rho \rangle$, has an obvious property that $\langle \delta\rho(\varepsilon) \rangle = 0$ and its equivalent

$$\int \delta\rho(\varepsilon)d\varepsilon = 0 \quad (2)$$

where the integral is taken over the spectrum.

Spectral correlations are characterized by the two-point correlation function

$$K(\varepsilon_1, \varepsilon_2) = \langle \delta\rho(\varepsilon_1)\delta\rho(\varepsilon_2) \rangle \qquad (3)$$

It immediately follows from Eq. (2) that

$$\int K(\varepsilon_1, \varepsilon_2) d\varepsilon_1 = \int K(\varepsilon_1, \varepsilon_2) d\varepsilon_2 = 0 \qquad (4)$$

In cases when the level repulsion sets in on a scale that is much smaller than $\varepsilon \gg 1$ (specifically, $\sim 1$ for chaotic systems and $\sqrt{\varepsilon}$ for integrable systems), the correlation function can be written as

$$K(\varepsilon_1, \varepsilon_2) = K(\varepsilon, \omega) \qquad (5)$$

with

$$\varepsilon = \frac{\varepsilon_1 + \varepsilon_2}{2}, \quad \omega = \varepsilon_1 - \varepsilon_2 \qquad (6)$$

Consequently, we also have

$$\int K(\varepsilon, \omega) d\omega = 0 \qquad (7)$$

The correlation function can further be separated into two terms [12]:

$$K(\varepsilon, \omega) = \delta(\omega) - \mathcal{K}(\varepsilon, \omega) \qquad (8)$$

where the first term corresponds to uncorrelated levels and the second term, which satisfies

$$\int \mathcal{K}(\varepsilon, \omega) d\omega = 1 \qquad (9)$$

corresponds to level repulsion.

As we pointed out, derivation of $K(\varepsilon, \omega)$ from the periodic orbit theory [2], [4] and classical dynamics is a challenging problem for chaotic systems. The main reason for it is a breakdown of the "diagonal approximation," in which interference between long periodic orbits is neglected. On the other hand, from RMT [1] and the non-linear model [9], the form of $\mathcal{K}(\varepsilon, \omega)$ is well known. For the unitary ensemble, for instance, it is given by

$$\mathcal{K}(\varepsilon,\omega) = \frac{\sin^2 \omega}{\pi \omega^2} \qquad (10)$$

which, as is the case for orthogonal and symplectic ensembles, is $\varepsilon$-independent and indicates that level repulsion develops on the scale of mean level spacing.

Our main interest is the form of $K(\varepsilon,\omega)$ for integrable systems. Unlike chaotic systems, it does not reduce to a simple closed form. On the other hand, the diagonal approximation is valid, which allows us to obtain the correlation function from the periodic orbits theory for simple classical dynamics. Furthermore, in such cases, it is also straightforward to obtain $K(\varepsilon,\omega)$ directly from the knowledge of the quantum mechanical spectrum.

In the periodic orbit theory, the Fourier transform of $K(\varepsilon,\omega)$ is given by [4] (in what follows, we set $\hbar = 1$)

$$K(\varepsilon,t) = \sum_j A_j^2 \delta(t - T_j) \qquad (11)$$

where $A_j$ and $T_j$ are amplitudes and periods of the periodic orbits at energy $\varepsilon$ [4] and it is known that

$$K(\varepsilon,\infty) = \frac{1}{2\pi} \qquad (12)$$

Previously, we introduced a simplified ansatz

$$K(\varepsilon,t) = \begin{matrix} 0 & t < T_{min} \\ 1/2\pi & T > T_{min} \end{matrix} \qquad (13)$$

that contains the key element of level repulsion. Here is the period of the shortest periodic orbit. Consequently we find [11]

$$K(\varepsilon,\omega) = \delta(\omega) - \frac{\sin(\omega T_{min})}{\pi \omega} \qquad (14)$$

The common feature (which is trivially obvious for hard wall billiards) of integrable systems with no extra symmetries is that

$$T_{min}^{-1} \propto \sqrt{\varepsilon} \gg 1 \qquad (15)$$

that is, level repulsion develops on a scale much larger than the mean level spacing. Nonetheless, it immediately affects the nearest level spacing distribution function via a relationship [15]

$$p(s) = g(s)\exp\left[-\int_0^s g(x)dx\right] \qquad (16)$$

where

$$g(x) = 1 - \mathcal{K}(x,\omega) \qquad (17)$$

Until now, $\mathcal{K}(x,\omega)$ had not been taken into consideration for integrable systems and the Poissonian distribution

$$p_P(s) = \exp(-s) \qquad (18)$$

was obtained as a result. With the simplified ansatz, the distribution becomes

$$p(s) = \left[1 - \frac{\sin(sT_{min})}{\pi s}\right]\exp\left[-s + \frac{\text{Si}(sT_{min})}{\pi}\right] \qquad (19)$$

and, in particular,

$$p(0) \approx 1 - \frac{T_{min}}{\pi} \qquad (19')$$

where Si is the sine integral. This is the central result of the paper. As will be shown below, it is convincingly confirmed numerically for the rectangular hard wall billiard and a modified Kepler problem discussed in the next Chapter.

## III. Rectangular billiard and modified Kepler Problem.

The level correlation can be explicitly evaluated for the rectangular billiard and the modified Kepler problem.[12]-[14] They can be also studied numerically against the analytical results obtained using the correlation function.

We begin with the rectangular hard wall billiards.[12] Without Balian-Bloch-like corrections [16], the level correlation function is found as

$$K(\varepsilon,\omega) = \frac{1}{\sqrt{\pi^3 \varepsilon}} \sum_{M_1,M_2=0}^{\infty} 4\delta_M \frac{\cos\left[\sqrt{4\pi/\varepsilon\left(M_1^2 \alpha^{1/2} + M_2^2 \alpha^{-1/2}\right)}\omega\right]}{\sqrt{\left(M_1^2 \alpha^{1/2} + M_2^2 \alpha^{-1/2}\right)}} \quad (20)$$

where $\alpha$ is the aspect ratio of the rectangle (assumed to be close to unity, $\alpha \sim 1$) and

$$\delta_M = \begin{cases} 0, & \text{if } M_1 = M_2 = 0 \\ 1/4, & \text{if one of } M_1 \text{ and } M_2 \text{ is zero} \\ 1, & \text{otherwise} \end{cases} \quad (21)$$

For the modified Kepler problem, described by the potential

$$V(r) = -\frac{\alpha}{r} + \frac{\beta}{r^2} \quad (22)$$

the relevant spectrum (in dimensionless units) is given by [13]

$$\varepsilon_{p,l} = 2p\sqrt{2\beta} + l^2 \quad (23)$$

for which the level correlation function is given by [14]

$$K(\varepsilon,\omega) = \sum_{M_r=1}^{\infty} \left(\left\lfloor \frac{M_r}{(2\beta/3\varepsilon)^{1/3}} \right\rfloor + \frac{1}{4}\right) \frac{2\sqrt{2\beta}}{\pi^2 M_r^3} \sin\left(\frac{\pi M_r \omega}{\left(3\varepsilon\sqrt{2\beta}\right)^{1/3}}\right) \quad (24)$$

where $M_r$ is the radial winding number and $\lfloor \ \rfloor$ is the floor function.

As explained in [11]-[14], averaging in equations (20) and (24) is understood as *parametric averaging*, namely averaging over the aspect ratio $\alpha$ for rectangles of the same area and over $\beta$ for the modified Kepler problem. This allows to sample ensembles with the same running energy, which is different from averaging over the energy spectrum that "washes away" important features of these systems, such as oscillations of the level number variance.

Persistent oscillations of the level number variance follows directly from (20) and (24) and are in excellent agreement with numerical simulations for both systems. [12],[14] Averaging over oscillations gives a "saturation value," which is twice that of the saturation rigidity. [4] Existence of these saturation values is directly related to the deviation of the nearest level distribution from Poissonian statistics and level repulsion in integrable systems.

Both (20) and (24) reduce to the form given by (14) for not too large $\omega$, up to $\sim T_{\min}^{-1}$. [12],[13] For brevity and in order to concentrate on the main result, in what follows we limit our consideration to the rectangular billiards. Specifically, we find

$$T_{\min} = \frac{2\pi^{3/2}}{\sqrt{\varepsilon}} \quad (25)$$

which, after substitution into (19), can be tested numerically.

### IV. Numerical Results.

We show results for the running energy (in units of mean level spacing, as explained above) $\varepsilon = 10^4$ obtained with $3 \times 10^5$ rectangular billiards, whose $3 \times 10^5$ values of aspect ratio $\alpha$ are from a normal distribution centered at 1 with half width of 0.2.

In Fig. 1, the red line is the Poissonian distribution; the dotted line is the numerical calculation; the green line is given by (19) with $T_{\min} = 0.11$ from (25); the purple line is the best fit obtained with (19), which yields $T_{\min} \approx 0.097$.

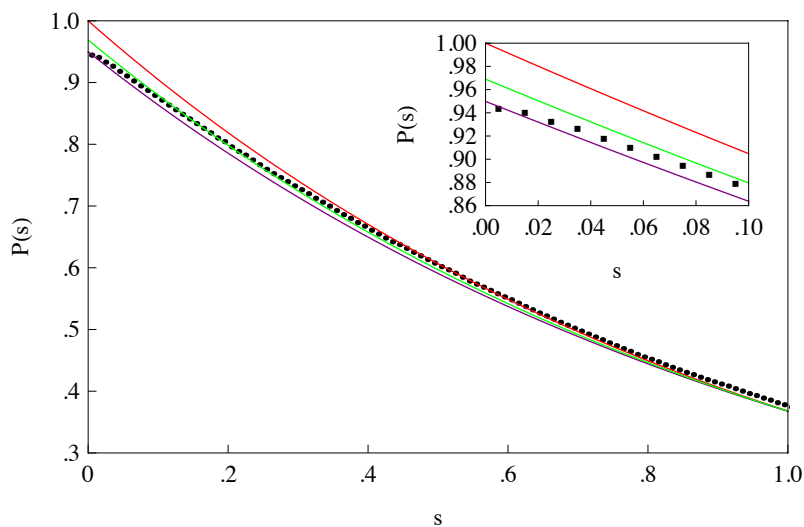

Fig. 1: Poissonian distribution (18) (red line), numerical calculation (dotted line) and distribution (19) (green and purple line – see text)

Deviation from Poissonian statistics is quite striking and is in good agreement with a simplified ansatz (19).

In Fig. 2, for a value of level separation $s \ll 1$, we investigate the following quantity:

$$P(s) = \frac{1}{s}\int_0^s p(s)ds \qquad (26)$$

as a function of running energy $\varepsilon$. For Poissonian statistics, this trivially yields

$$P_P(s) = \frac{1-\exp(s)}{s} \qquad (27)$$

which is $\varepsilon$-independent. Using now (19) in (26), we find

$$P(s) = \frac{1-\exp[-s+\mathrm{Si}(sT_{\min})/\pi]}{s} \stackrel{s \ll 1}{\approx} \frac{1-\exp(-s)}{s} - \frac{T_{\min}}{\pi} = P_P(s) - 2\sqrt{\frac{\pi}{\varepsilon}} \qquad (28)$$

and, choosing a value of level separation $s = 0.05$, we compare the Poissonian value $P_P(s=0.05) = 0.976$ (red line) to the expression given by (28). As in Fig. 1, the dotted line is the numerical calculation, the green line is (28), $P_P(s) - 3.54/\sqrt{\varepsilon}$; the purple line is the best parametric fit obtained with (28), $P_P(s) - 3.19/\sqrt{\varepsilon}$.

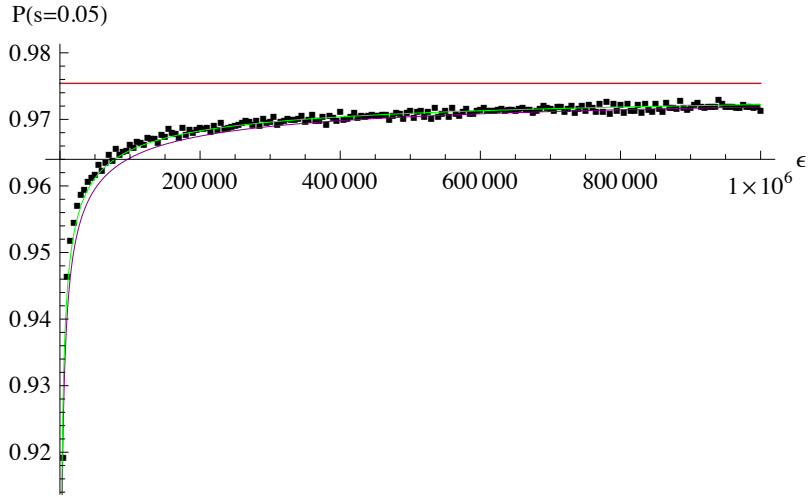

Fig. 2: Poissonian distribution (27) (red line), numerical calculation (dotted line) and (28) (green and purple line – see text)

This provides an even more striking demonstration of the level repulsion and of the deviation from the Poissonian statistics. We obtained similarly convincing results (with somewhat more challenging numerical calculation) for the modified Kepler problem (22)-(24), which will be presented elsewhere.

**Conclusions**

We have demonstrated existence of level repulsion in the semiclassical spectrum of classically integrable systems. This phenomenon is a direct consequence of strong level correlations as expressed by equations (20) and (24). These correlations are also responsible for the very unusual behavior of the level rigidity and level number variance. [12]-[14] We can account for deviations from the Poissonian statistics using a simplified ansatz (14), which results in a distribution given by (19). Figs. 1 and 2 give a compelling numerical support to our findings.